\documentclass[a4paper,showpacs,twocolumn,amsmath,amssymb,floatfix,prb,preprintnumbers,footinbib]{revtex4}
\usepackage{graphicx}
\usepackage{amsmath,amsfonts}
\usepackage{color}

\newcommand{\e}{\varepsilon}
\newcommand{\m}[1]{\mathrm{#1}}

\begin{document}

\title{Spin-dependent transport through quantum-dot Aharonov-Bohm interferometers}

\author{Bastian Hiltscher$^1$, Michele Governale$^2$, and J\"urgen K\"onig$^1$} 
\affiliation{$^1$Theoretische Physik, Universit\"at Duisburg-Essen and CeNIDE, 47048 Duisburg, Germany\\
$^2$School of Chemical and Physical Sciences  and MacDiarmid Institute for Advanced Materials and Nanotechnology, Victoria University of Wellington, P.O. Box 600, Wellington 6140, New Zealand}

\date{\today}
\begin{abstract}
We study the influence of spin polarization on the degree of coherence of electron transport through interacting quantum dots. 
To this end, we identify transport regimes in which the degree of coherence can be related to the visibility of the Aharonov-Bohm oscillations in the current through a quantum-dot Aharonov-Bohm interferometer with one normal and one ferromagnetic lead. 
For these regimes, we calculate the visibility and, thus, the degree of coherence, as a function of the degree of spin polarization of the ferromagnetic lead.
\end{abstract}

\pacs{73.23.Hk, 85.35.Ds, 72.10.Bg}

\maketitle
\section{Introduction}

The investigation of electric transport properties of nano-structured devices defines a field of increasing importance. 
The issue of quantum coherence and its limitation by Coulomb interaction can be conveniently studied in devices that contain quantum dots (QDs) in multiply-connected geometries.
The interplay between interference and Coulomb interaction has been extensively studied in these so-called quantum-dot Aharonov-Bohm interferometers (QD-ABIs) both experimentally\cite{aikawa04,ihn07,yacoby94,schuster97,ji00,wiel00,holleitner01,sigrist06,gustavsson08} and theoretically.~\cite{aleiner97,khym06,moldo07,hofstetter01,koenig01,koenig02,akera93,yeyati95,bruder96,hackenbroich96,wu98,kang99,silvestrov00,gerland00,kang00,boese01,silvestrov03, lopez05,urban08,silva09}
Observed oscillations of the current through quantum-dot ABIs as a function of the magnetic flux enclosed
by the interferometer arms~\cite{yacoby94,schuster97,ji00,wiel00,holleitner01,aikawa04,sigrist06,gustavsson08,ihn07} 
prove that transport through a quantum dot is at least partially coherent.
The degree of coherence may be suppressed by interaction. 
This can, e.g., be studied in a controlled way by electrostatically coupling a quantum-point contact (QPC) to the quantum dot in the ABI.
The current through the QPC serves as a which-path detector that diminishes the amplitude of the interference signal.~\cite{aleiner97,moldo07,khym06,gustavsson08}

But even in the absence of any coupling to the outside world the degree of coherence may be limited by Coulomb interaction among the electrons within the QD-ABI.
This is the issue that we will concentrate on for the rest of the paper.
Similarly, the effect of different interdot- and intradot-interactions in T-shaped quantum-dot interferometers on the amplitude of the Fano resonance has been studied.\cite{moldoveanu08}
The two central questions that we will address are: (1) What fraction $c=I^{\rm coh}/I^{\rm total}$ of the total current through a single-level quantum dot weakly coupled to the electrodes is coherent? (2) How and under which circumstances can this fraction $c$ be extracted from a current measurement in an Aharonov-Bohm setup?
For transport through a single-level quantum dot with strong Coulomb interaction weakly coupled to normal leads the answer was given in Refs.~\onlinecite{koenig01} and \onlinecite{koenig02}. If any coupling of the quantum dot to some bath is negligibly small then the only source of decoherence is connected to the spin degree of freedom in the quantum dot. 
In general, transport through the quantum dot can be divided into spin-flip and non-spin-flip processes. Spin-flip processes due to spin-orbit coupling are neglected in the following considerations. 
Furthermore, we restrict our analysis to temperatures larger than the Kondo temperature.\cite{remark}
When the dot is initially empty, transferred electrons keep their spin orientation, and the transport is fully coherent. 
In contrast, when the dot is occupied with a single electron, then the transferred electron may either keep or flip its spin, i.e., only half of the processes (the non-spin-flip ones) are coherent.
As a result, the fraction of coherent to total linear conductance in the limit of weak tunnel coupling is $c=1/[1+f(\epsilon)]$, where $f(\epsilon)$ is the Fermi function and $\epsilon$ the quantum dot level, measured relative to the Fermi energy of the leads.
It was theoretically predicted\cite{koenig01,koenig02} and experimentally confirmed\cite{aikawa04,ihn07} that this fraction $c$ of coherent transport can be extracted from measuring the Aharonov-Bohm oscillation amplitude as a function of level energy for a quantum dot embedded in an Aharonov-Bohm ring.
The asymmetry of the oscillation amplitude for $\epsilon>0$ as compared to the one for $\epsilon<0$ was in agreement with the theoretical prediction. The restriction to a single level is justified as long as the level spacing on the dot is larger than temperature and bias voltage such that only one orbital participates in transport. The influence of many levels on coherence and the crossover from large to small level spacing is discussed in Ref. \onlinecite{koenig02}.

In order to substantiate the role played by the spin, we suggest in this paper to replace one of the electrodes by a lead with a finite degree of spin polarization $p$.
The main idea behind this proposal is that a large degree of spin polarization should, in general, increase the fraction of coherent transport since spin-flip processes are less frequent.
However, introducing a spin-polarized lead breaks the spin symmetry and, thus, changes the transport characteristics in a non-trivial way.
This includes the possibility of spin accumulation on the dot,\cite{linde09,braun04,souza07,barnas08}, tunnel magneto resistance\cite{stefanski09,hamaya07} or a negative differential conductance.\cite{braun04,bulka00,rudzinski05,braig05}
Therefore, both questions 1. and 2. have to be reanalyzed carefully.
Since the physics of spin accumulation may introduce an asymmetry of the current between the cases $\epsilon>0$ and $\epsilon<0$, that is not related to decoherence, an asymmetry of the AB oscillation amplitude does not necessarily indicate decoherence. 
The measurable quantity to compare $c$ with is the visibility $v$.
In case of weak tunneling, where only one Fourier component of the flux-dependent current needs to be considered, the visibility $v$ (with $v>0$) is defined via
\begin{equation}
	I^{\rm total}(\varphi) = I^{\rm av} \left[ 1 + v  \cos(\varphi +\delta) \right]\, ,
\end{equation}
where $I^{\rm av}$ is the flux-averaged current and $\varphi$ the AB phase.
As we will argue below, a clear correspondence between $c$ and $v$ can be established in the regime of uni-directional cotunneling, with an extra condition for the polarity of the applied bias voltage in the case $\epsilon<0$.
In the latter case, it is possible to extract a polarization-dependent coherence factor $c=(1+p^2)/2$ by measuring the visibility $v$.

This paper is structured as follows. In Section \ref{model} we introduce the model under consideration. The theoretical method that we employ is described in Sec. \ref{method}. Results are presented in Sec. \ref{results}, which is subdivided in four parts: expressions for the charge current in different orders are given in  Sec. \ref{current}; we discuss the fraction of coherent transport in Sec.~\ref{rescoherence} ; Sec. \ref{secfluxdep} concerns the visibility of the current and Sec. \ref{seccoherence} elucidates the relation between visibility and coherence. Conclusions are drawn in Sec. \ref{conclusions}.

\section{Model}
\label{model}
\begin{figure}[h]
\centering
\includegraphics[width=7cm]{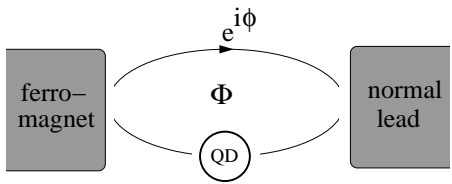}
\caption{\label{setup}Setup of single-dot Aharonov-Bohm interferometer with one spin-polarized lead.
}
\end{figure}

We consider a closed single-dot ABI, i.e., a two-terminal ABI with a
single-level quantum dot embedded in one of the arms, see Fig \ref{setup}. The total Hamiltonian of our system consists of four parts,
\begin{eqnarray}
H=H_{\text{dot}}+H_{\text{leads}}+H_{\text{tunn}}+ H_{\text{ref}}\, .
\end{eqnarray}

The QD is assumed to accommodate a single, spin-degenerate level. It is described by the Anderson-impurity model: 
\begin{eqnarray}\label{dothamil}Ê 
H_{\text{dot}}&=&\e\sum\limits_\sigma n_{\sigma} + Un_{\uparrow} n_{\downarrow \,} .
\end{eqnarray}
Here, $n_{{\sigma}}=d_{\sigma}^\dagger d_{\sigma}$, with $d_{\sigma}^\dagger$ being the creation operator for an electron with spin $\sigma$ on the quantum dot.  The dot-level position is denoted by $\epsilon$ and the onsite Coulomb-repulsion energy by $U$. In the results, we will concentrate on the two limits of non-interacting electrons, $U=0$, and infinite charging energy, $U=\infty$.

We consider a two terminal setup, with the index $r=\text{F}$ labeling the ferromagnet and $r=\m{N}$ labeling the normal conductor. Both leads 
are large, non-interacting reservoirs, whose Hamiltonian reads 
\begin{eqnarray}
H_{\text{leads}}&=&\sum\limits_{r k\sigma}\e_{rk\sigma}c_{rk\sigma }^\dagger c_{rk\sigma
}\, ,
\end{eqnarray}
where  $ c_{r k\sigma}^\dagger$ is the creation operator for an electron in lead $r$ in a state labeled by the quantum number $k$ and with spin $\sigma$. 

The tunnel coupling between the dot and the two leads is modeled by the tunneling Hamiltonian:
\begin{eqnarray}\label{tunnelhamil}
H_{\rm{tunn}}&=&\sum\limits_{r k\sigma }t_{r}  c_{r k\sigma}^\dagger d_{\sigma}
+\mathrm{H.c.} \, .
\end{eqnarray}
We  assume the tunnel matrix elements $t_{r}$ and the density of states of the leads $N_{r\sigma}$ to be energy independent in the energy window relevant for transport. In the ferromagnetic lead we also have to distinguish between the density of states of electrons with majority ($\sigma=+$) and minority spin ($\sigma=-$). For the normal lead this distinction is not necessary ($N_\m{N}/2\equiv N_{\m{N}+}= N_{\m{N}-}$).  The spin polarization $p=(N_{\m{F}+}-N_{\m{F}-})/(N_{\m{F}+}+N_{\m{F}-})$ characterizes the asymmetry of the density of states. Tunnel-coupling strengths are then defined as $\Gamma_\m{N}=2\pi |t_\m{N}|^2 N_\m{N}$ and  $\Gamma_\m{F \pm}=2\pi |t_\m{F}|^2 N_\m{F\pm}= (1\pm p)\Gamma_\m{F}$.
The intrinsic line width of the quantum dot's level is the sum of the tunnel couplings,
 $\Gamma=\Gamma_\m{N}+\Gamma_\m{F}$.
 
The second (``reference'') interferometer arm is modeled by a direct tunnel coupling between the leads. The Hamiltonian of the reference arm reads
 \begin{eqnarray}
H_{\text{ref}}=\sum\limits_{k \in \m{N}, q \in \m{F}, \sigma} (\tilde{t}c^\dagger_{\text{N} k\sigma}c_{\text{F} q\sigma }+\m{H.c.})\, .
\end{eqnarray}
with transmission amplitude $t^{\text{ref}}_\sigma=2\pi \tilde{t} \sqrt{N_{\text{F}\sigma}N_{\text{N}}}$. The magnetic flux $\Phi$  threading the interferometer is included in the phases of the tunneling amplitudes. We choose the gauge in which $t_{\text{F}},t_{\text{N}} \in \Re^+$ and $\arg \tilde{t}=\varphi=2\pi \Phi/\Phi_0$, where $\Phi$ is the magnetic flux and $\Phi_0$ the flux quantum.
In analogy to the tunnel coupling to the dot, we define the total transmission probability 
$|t^{\text{ref}}|^2= |t^{\text{ref}}_+|^2 + |t^{\text{ref}}_-|^2$.

\section{Method}
\label{method}

The dynamics of the quantum dot's degree of freedom, i.e., the probabilities $P_{\chi}$ to find the dot in state $\chi=0, \uparrow, \downarrow, d$, is governed by a generalized master equation.
In the stationary limit, it reads $0=\sum_{\chi'}W_{\chi\chi'}P_{\chi'}$, where $W_{\chi\chi'}$ are the transition rates from state $\chi'$ to $\chi$.
Having solved the master equation for the probabilities, the stationary current can be computed from $I=\m{e}\sum_{\chi'}W_{\chi\chi'}^IP_{\chi'}$, where the current transition rates $W_{\chi\chi'}^I$ are obtained from the transition rates $W_{\chi\chi'}$ by multiplying with the net number of electrons that are transfered from source to drain in the transition described by $W_{\chi\chi'}$.

Our method is applicable for arbitrary values of the Coulomb repulsion $U$. However, for simplicity we only consider the two limits $U=0$ and $U=\infty$ from now on. In the latter case, double occupancy of the dot is prohibited.  
We aim at a systematic perturbation expansion for weak coupling ($\Gamma\lesssim k_\m{B}T$ and $|t^\m{ref}|\ll1$) of the current $I=\sum_{m,n} I^{(m,n)}$, where $m$ indicates the power in the tunnel coupling $\Gamma$ between dot and leads  and $n$ the power in the direct tunnel coupling $|t^\m{ref}|$ between the two leads. 
A direct coupling between the leads can be made small in experiments with the help of a tunable barrier in the reference arm. We perform a corresponding expansion for the probabilities and the transition rates.
We restrict ourselves to the lowest-order contributions.
This means, we include the current through the reference arm in the absence of the quantum dot, $I^\m{(0,2)}$, the interference term $I^\m{(1,1)}(\varphi)$, which is the lowest-order contribution that depends on the Aharonov-Bohm phase $\varphi$, and the current through the quantum dot in the absence of the reference arm.

For the last contribution, it is important to distinguish two different transport regimes.
If the dot level $\e$ lies inside the energy window for which occupied states in the source electrode and simultaneously empty states in the drain are available, i.e., 
$|\e| \lesssim \max\{k_\m{B}T,|\m{e}V/2|\}$, then transport is dominated by transition rates 
$W^\m{(1,0)}$ (and $W^{I\m{(1,0)}}$) that are first order in $\Gamma$.
It is clear that in this case only first-order rates are required to evaluate the zeroth-order probability distribution $P_\chi^{(0,0)}$.
We refer to this procedure as calculation scheme 1. 

The situation is different in the cotunneling regime, $|\e| \gg \max\{k_\m{B}T,|\m{e}V/2|\}$, for which some of the rates $W^\m{(1,0)}$ are exponentially suppressed and the lowest-order contribution is $W^\m{(2,0)}$.
Then, as discussed e.g. in Ref.~\onlinecite{weymann05/2}, some second-order rates are required to evaluate the zeroth-order probability distribution $P_\chi^{(0,0)}$. This we call calculation scheme 2.

For scheme 1, we use a real-time diagrammatic technique to perform the perturbation expansion in the tunnel-coupling strengths.\cite{koenig96} 
The advantage of this technique is that it is systematic in the sense that all contributions of given order are properly taken into account. The downside is that including higher-order contributions becomes increasingly cumbersome.
In the cotunneling regime, where scheme 2 needs to be used, the expressions for the rates obtained from the diagrammatic technique drastically simplify.
In that case it is easier to directly identify all the cotunneling processes and evaluate the corresponding rates by second-order perturbation theory rather than employing the real-time diagrammatics. 

To discuss the results obtained by scheme 1, we will only provide the final expressions for the current. 
As we will discuss below, for connecting the degree of coherence with the visibility of the Aharonov-Bohm oscillations, the cotunneling regime is more important.
In this case, we use scheme 2 with the cotunneling rates
$W^{(2,0)}_\m{\chi\chi'}=\sum\limits_{r,r'} \gamma^{\chi\chi'(2,0)}_{rr'}$, where
$\gamma^{\chi\chi'}_{rr'}$ is the rate of a transition where an electron is transfered from reservoir $r'$ to reservoir $r$, 
accompanied by a change of the dot state from $\chi'$ to $\chi$. 
An example for the calculation of such a cotunneling rate, as introduced in 
Refs.~\onlinecite{averin89,averin90} for metallic islands and applied for single-level quantum dots, e.g., in 
Ref.~\onlinecite{weymann05/1}, is given in Appendix \ref{secordcalc}.
For $|\e| \gg \max\{k_\m{B}T,|\m{e}V/2|\}$ and $U=0$, the cotunneling rates simplify to
\begin{eqnarray}
	\gamma^{00(2,0)}_{rr'}&=&\gamma^{\sigma\sigma(2,0)}_{rr'}=\gamma^{dd(2,0)}_{rr'}=
	\sum_{\sigma'} \frac{\Gamma^{\sigma'}_{r}\Gamma^{\sigma'}_{r'}}{2\pi\e^2} 
	F(\mu_r-\mu_{r'}) 
	\\
	\gamma^{\bar{\sigma}\sigma(2,0)}_{rr'}&=&0
\end{eqnarray}
with $F(x)=x/[\exp (x/k_\m{B}T)-1]$. 
For $U=\infty$, we obtain
\begin{eqnarray}
	\gamma^{00(2,0)}_{rr'}&=&\sum_\sigma \frac{\Gamma^\sigma_{r}\Gamma^\sigma_{r'}}{2\pi\e^2} 
	F(\mu_r-\mu_{r'}) 
	\label{rate_empty}
	\\
	\gamma^{\sigma\sigma(2,0)}_{rr'}&=&\frac{\Gamma^\sigma_{r}\Gamma^\sigma_{r'}}{2\pi\e^2} 	F(\mu_r-\mu_{r'}) 
	\label{rate_nonsf}
	\\
	\label{rate_sf}
	\gamma^{\bar{\sigma}\sigma(2,0)}_{rr'}&=&\frac{\Gamma^{\sigma}_{r}\Gamma^{\bar{\sigma}}_{r'}}{2\pi\e^2}   F(\mu_r-\mu_{r'}) 
	\, .
\end{eqnarray}
The rates in Eqs.~(\ref{rate_empty}) and (\ref{rate_nonsf}) are associated with non-spin-flip processes while Eq.~(\ref{rate_sf}) describes spin-flip processes.

Finally, we also need the rates to first order in $\Gamma$ and first order in $|t^\m{ref}|$.
Only those contributions with $r \neq r'$ exist.
We obtain 
\begin{eqnarray}
	\gamma^{00(1,1)}_{rr'}&=& \delta_{r,\bar r'}\sum_\sigma \frac{|t^\m{ref}_\sigma| \sqrt{\Gamma^\sigma_{r}\Gamma^\sigma_{r'}}}{\pi\e}F(\mu_r-\mu_{r'}) \cos\varphi 
\\
	\gamma^{\sigma\sigma'(1,1)}_{rr'}&=& 2 \delta_{\sigma,\sigma'} \delta_{r,\bar r'} \frac{|t^\m{ref}_\sigma| \sqrt{\Gamma^\sigma_{r}\Gamma^\sigma_{r'}}}{\pi\e}F(\mu_r-\mu_{r'}) \cos\varphi 
\\
	\gamma^{dd(1,1)}_{rr'}&=&- \gamma^{00(1,1)}_{rr'}
\end{eqnarray}
for $U=0$ and
\begin{eqnarray}
	\gamma^{00(1,1)}_{rr'}&=& \delta_{r,\bar r'}\sum_\sigma \frac{|t^\m{ref}_\sigma| \sqrt{\Gamma^\sigma_{r}\Gamma^\sigma_{r'}}}{\pi\e}F(\mu_r-\mu_{r'}) \cos\varphi 
\\
	\gamma^{\sigma\sigma'(1,1)}_{rr'}&=& \delta_{r,\bar r'} \frac{|t^\m{ref}_\sigma| \sqrt{\Gamma^\sigma_{r}\Gamma^\sigma_{r'}}}{\pi\e}F(\mu_r-\mu_{r'}) \cos\varphi 
\end{eqnarray}
for $U=\infty$. 
Here, $\bar r'$ indicates the lead other than $r'$. 
 
\section{Results}
\label{results}

\subsection{Charge current}
\label{current}

The quantity that is directly measured in experiment is the charge current.
As indicated above, the total current can be split into three contributions: the current through the reference arm in the absence of the quantum dot, $I^\m{(0,2)}$, the current through the quantum dot in the absence of the reference arm, $I^\m{(1,0)}$ or $I^\m{(2,0)}$ (for scheme 1 and 2, depending on the level position $\epsilon$, respectively), and the interference term, $I^\m{(1,1)}(\varphi)$. 
Only the last one depends on the Aharonov-Bohm phase $\varphi$. 

Direct tunneling through the reference arm can be calculated with Fermi's golden rule and contributes to the current with
\begin{equation}
	I^\m{(0,2)}=\frac{e^2}{\pi}V|t^\m{ref}|^2,
\end{equation}
where $V$ is the bias voltage applied between the ferromagnet and the normal conductor.

We now consider the transport through the quantum dot in the absence of the direct interferometer arm.
If the dot level lies in between the transport voltage defined by the Fermi energies of the electrodes then transport through the dot will be dominated by first-order tunneling, $I^\m{(1,0)}$, and we use scheme 1.
We find 
\begin{eqnarray}
	I^\m{(1,0)}=-2\m{e}\frac{\Gamma_\m{F}\Gamma_\m{N}\left(\Gamma-p^2\Gamma_\m{F} \right)}	{\Gamma^2-p^2\Gamma^2_\m{F}}\left[f_\m{F}(\e)-f_\m{N}(\e)\right]\, ,
\end{eqnarray}
for noninteracting electrons, $U=0$, where $f_\m{F/N}$ is the Fermi function of the normal/ferromagnetic lead. For an infinite interaction, $U=\infty$, we obtain
\begin{eqnarray}
	I^\m{(1,0)}=-2\m{e}A^{-1}\Gamma_\m{F}\Gamma_\m{N}\left[f_\m{F}(\e)-f_\m{N}(\e)\right]\nonumber \\ 
	\times\left[\Gamma_\m{F}\left(1-p^2\right)(1-f_\m{F}(\e))+\Gamma_\m{N}\left(1-f_\m{N}(\e)\right)\right]     \, ,
\end{eqnarray}
with
\begin{eqnarray}
	A=\Gamma^2-p^2\Gamma_\m{F}^2-[\Gamma_\m{F}f_\m{F}(\e)+ \Gamma_\m{N}
	f_\m{N}(\e)]^2+p^2\Gamma_\m{F}^2f_\m{F}^2(\e) \nonumber  \, .
\end{eqnarray}

If the dot level lies outside the energy window defined by the Fermi energies of the leads  ($|\e|\gg\max\{k_\m{B}T,|\m{e}V/2|\}$) then 
$I^\m{(1,0)}$ is exponentially suppressed and transport through the dot is dominated by cotunneling, 
$I^\m{(2,0)}$. In this case, we employ scheme 2, see Sec. \ref{method}.
In this regime the current for noninteracting electrons ($U=0$) reads,
\begin{equation}
	I^\m{(2,0)}=\m{e}^2\frac{\Gamma_\m{F}\Gamma_\m{N}}{\pi\e^2}V\, .
\end{equation}

In the case of an infinite Coulomb interaction on the dot ($U=\infty$) we have to distinguish different cases. For a dot-level position well above the Fermi energy of the leads, $\e>0$, the current through the quantum dot is the same as for noninteracting electrons. 
In the opposite case, $\e<0$, we get
\begin{equation}
	I^\m{(2,0)}=\m{e}^2\frac{\Gamma_\m{F}\Gamma_\m{N}}{\pi\e^2}V\left[1+\frac{pm}{1-\m{exp}(-\m{e}V/k_\m{B}T)}\right]\, ,
\end{equation}
where $m$ is the spin accumulation on the dot, which depends on the transport direction. In the regime of unidirectional cotunneling, $|\e|\gg|\m{e}V/2|\gg k_\m{B}T$, it simplifies to $m=p$ for transport from the ferromagnetic into the normal lead ($V<0$) and $m=-p$ for the opposite transport direction ($V>0$).

The flux-dependent part is given by $I^{(1,1)}(\varphi) = I^{(1,1)}_\m{even}\cos \varphi + I^{(1,1)}_\m{odd}\sin \varphi$.
For noninteracting electrons, the coefficients are 
\begin{eqnarray}
	I^\m{(1,1)}_\m{odd}=0 
\end{eqnarray}
and
\begin{eqnarray}
	I^\m{(1,1)}_\m{even}=2\m{e} |t^\m{ref}|\sqrt{\Gamma_\m{N}\Gamma_\m{F}}\sigma(\e) \,
\end{eqnarray}
with
\begin{eqnarray*}
\sigma(\e)=\frac{1}{\pi}\m{Re}\left[\int d\omega\frac{f_\m{F}(\omega)-f_\m{N}(\omega)}{\e-\omega+\m{i}0^+}\right]\, ,
\end{eqnarray*}
independent of the polarization $p$. 

For an infinitely strong charging energy, both the contributions even and odd in the flux are present.
They read in the sequential tunneling regime (scheme 1)
\begin{eqnarray}
	I^\m{(1,1)}_\m{odd}&=&2\m{e}A^{-2}|t^\m{ref}|(\Gamma_\m{N}\Gamma_\m{F})^{3/2}\left(f_	\m{F}(\e)-f_\m{N}(\e)\right)^2 \nonumber \\
	&\times& \left\{\left[\Gamma_\m{F}(1-p^2)(1-f_\m{F}(\e))+\Gamma_\m{N}(1-f_\m{N}(\e))\right]^2\right.\nonumber\\
	&-&\left.\Gamma_\m{N}^2\left[p^2\left(1-f_\m{N}^2(\e)\right)\right]\right\} \,
\end{eqnarray}
and
\begin{eqnarray}
	I^\m{(1,1)}_\m{even}&=&2\m{e}A^{-1}|t^\m{ref}|\sqrt{\Gamma_\m{N}\Gamma_\m{F}}	\sigma(\e) \nonumber \\
	&\times& \left\{\Gamma_\m{F}^2(1-f_\m{F}(\e))(1-p^2)+\Gamma_\m{N}^2(1-f_\m{N}(\e))\right.\\
	&+&\left.\Gamma_\m{F}\Gamma_\m{N}\left(2-f_\m{F}(\e)(1-p^2)-f_\m{N}(\e)(1+p^2)\right)\right\} \nonumber\,
\end{eqnarray}
respectively.

In the cotunneling regime (scheme 2) the odd contribution drops out, $I^\m{(1,1)}_\m{odd}=0$ while the even part is given by 
\begin{equation}
I^\m{(1,1)}_\m{even}=-\frac{\m{e}^2V}{\pi\e}\sqrt{\Gamma_\m{F}\Gamma_\m{N}}|t^\m{ref}|
(1+pm)\; .
\end{equation}

\begin{figure}[h]
\centering
\includegraphics[width=6cm]{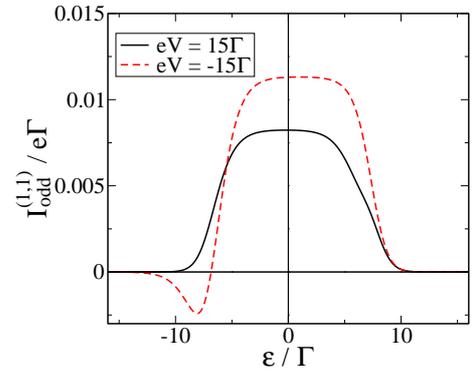}
\caption{\label{Iodd}(Color online) Odd part of the first flux dependent order of the current $I^\m{(1,1)}_\m{odd}$ for polarization $p=0.7$ and Coulomb interaction $U=\infty$ as a function of $\e$.  In the total current a negative bias voltage corresponds to a transport from the ferromagnet into the normal conductor. 
The value of the parameters used in the calculations are: $\varphi=\pi/2$, $|t^\m{ref}|=0.1$, $k_\m{B}T=\Gamma$, $\Gamma_\m{F}=\Gamma_\m{N}=\Gamma/2$.} 
\end{figure}

The odd and even parts of the first flux dependent correction differ in many respects. The odd part $I^\m{(1,1)}_\m{odd}$ describes transport processes where an electron cotunnels through a lead.\cite{urban08} It only occurs for a nonvanishing Coulomb interaction. Figure \ref{Iodd} shows  both transport directions of $I_\m{odd}^\m{(1,1)}$ as a function of the dot-level position $\e$ for an infinite Coulomb interaction. The current has its maximum value where the dot level lies between the chemical potential of the two leads. Beside this range the current decreases exponentially.

\begin{figure}[h]

\includegraphics[width=6cm]{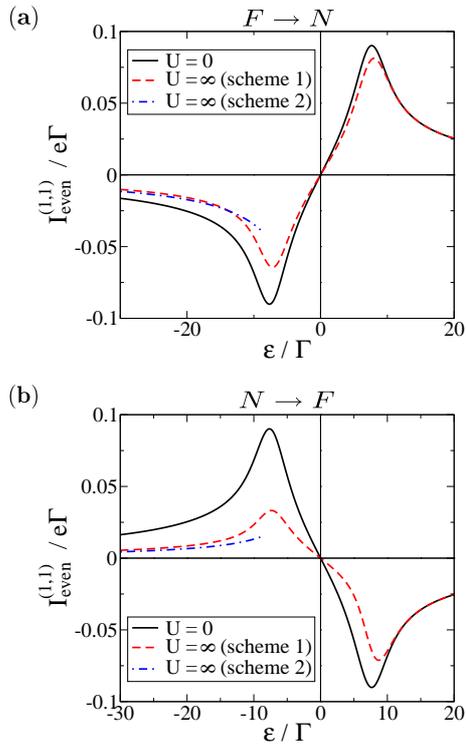}
\caption{\label{strom}(Color online) Even part of the first flux dependent order of the current $I^\m{(1,1)}_\m{even}$ for polarization $p=0.7$ as a function of $\e$ for vanishing (solid line) and infinite Coulomb interaction for two different calculation schemes (see text). Scheme 1 (dashed line) is more accurate for $\e\gtrsim -\max\{k_\m{B}T, |\m{e}V/2|\}$ while scheme 2 (dashed-dotted line) is more accurate for $\e\ll -\max\{k_\m{B}T, |\m{e}V/2|\}$. \textbf{a)} Electrons are transported from ferromagnet into normal lead ($eV=-15\Gamma$). \textbf{b)} Electrons are transported from normal lead into ferromagnet ($eV=15\Gamma$). The value of the parameters used in the calculations are: $\varphi=0$, $|t^\m{ref}|=0.1$, $k_\m{B}T=\Gamma$, $\Gamma_\m{F}=\Gamma_\m{N}=\Gamma/2$. } 
\end{figure}

Figure \ref{strom} shows $I_\m{even}^\m{(1,1)}$ as function of dot-level position for vanishing and infinite Coulomb interaction. For an infinite Coulomb interaction  two different lines are shown. The red, dashed line is calculated by means of scheme 1 while the blue, dashed-dotted line is obtained by means of scheme 2, see Sec. \ref{method}. Figure  \ref{strom}a) shows the current of electrons from the ferromagnetic into the normal lead. For noninteracting electrons the current $I^\m{(1,1)}$ is an odd function of the dot level position $\e$. The sign change around $\e=0$ relies on a phase shift of the transmission amplitude of the quantum dot. An infinite Coulomb interaction excludes the double occupation of the dot. Hence it has a higher influence for negative than for positive values of $\e$. The transport from the normal conductor into the ferromagnet is for $\e<0$ strongly suppressed, see Fig. \ref{strom}b). Transport through the quantum dot is blocked by an accumulation of the minority spin on the dot. 

What can we conclude from this for the fraction $c$ of coherent transport through a quantum dot?
Not much, as long as the dot's level is inside the energy window of lowest-order transport. 
And even for the cotunneling regime, an interpretation is difficult for $|\m{e}V/2|\lesssim k_\m{B}T$, i.e., when transport processes from source to drain are partially compensated by processes from drain to source.
For the further discussion, we will, therefore, turn to the regime of unidirectional cotunneling, $|\e|\gg|\m{e}V/2|\gg k_\m{B}T$. 
We emphasize that our method is applicable to arbitrary values of $U$. For $U=0$, no spin-flip processes occur since contributions with intermediate empty and double occupation of the dot cancel out each other. As long as $U\ll\min\{|eV|,k_\m{B}T\}$ this also holds for a finite $U$.  In the opposite limit, $U=\infty$, double occupancy is fully suppressed, and this cancellation does not occur anymore. This will remain true as long as $U\gg\max\{|eV|,k_\m{B}T\}$ . Between these two limits there will be a smooth crossover. Therefore, we focus on the limits $U=0$ and $U=\infty$ only. In particular we will distinguish the four different cases summarized in the Table~\ref{cases}. 

\begin{table}[h]\caption{\label{cases}The considered cases}
\begin{tabular}{|l|c|}
\hline
case 1&$U=0$\\ \hline
case 2a& $U=\infty$, $\e\gg|\m{e}V/2|\gg k_\m{B}T$
\\ \hline
case 2b&$U=\infty$, $-\e\gg|\m{e}V/2|\gg k_\m{B}T$, $F\rightarrow N$
\\ \hline
case 2c&$U=\infty$, $-\e\gg|\m{e}V/2|\gg k_\m{B}T$, $N\rightarrow F$
\\ \hline
\end{tabular}
\end{table}

For reference, we always compare to the non-interacting limit (case 1).
For strong Coulomb interaction, the dot level may either lie well above the Fermi level of the leads (case 2a), or it may lie well below.
In the latter case, the results will strongly depend on the polarity of the applied transport voltage. 
Case 2b refers to the limit when electrons are transported from the ferromagnet to the normal lead, and case 2c describes the opposite transport direction.

\subsection{Fraction of coherent transport}
\label{rescoherence}

How can the fraction of coherent transport be measured in an experiment?
Coherence can be tested by interferometry. 
We consider here an Aharonov-Bohm interferometer in which a single-level quantum dot is embedded in one of the arms. 
Electrons entering from the source electrode can either travel through the quantum dot or through the direct arm to the drain.
If no spin flip occurs, there will be an interference of both paths, which gives rise to a flux-dependent current.
The amplitude of the Aharonov-Bohm oscillations relative to the flux-averaged current contains information about the degree of coherence.

There is, however, a major problem in quantitatively connecting the degree of coherence $c$ and the visibility $v$. 
A quantum-dot Aharonov-Bohm interferometer probes many different transport channels, distinguished by the energy of the incoming electron, simultaneously.
If for the participating electrons the transmission through the quantum dot is strongly energy dependent, then the expected visibilities for the individual channels will be very different from each other.
This is the case, when the dot's level position lies inside the energy window defined by the Fermi energies of the leads.
For establishing a connection between visibility and fraction of coherent transport, we need to identify a situation in which for all participating electrons the transmission through the dot is the same. 
This is possible in the cotunneling regime, i.e., when the energy level of the quantum dot is outside this energy window, $|\e|\gg\max \{|\m{e}V/2|,k_\m{B}T\}$.
Furthermore, for the case when the fraction of coherent transport depends on the transport direction, we need $|\m{e}V| \gg k_\m{B}T$, i.e., uni-directional cotunneling, as an extra condition to separate the two directions. 

What do we expect for the fraction $c$ of coherent transport in this regime of uni-directional cotunneling from lead $r$ to lead $\bar r$?
We assume that flipping the spin in the quantum dot provides the only source of decoherence.
The coherence fraction $c$ is the ratio 
\begin{equation}
	c=\frac{\sum_{\chi} \gamma^{\chi\chi(2,0)}_{\bar{r}r}P_{\chi}^\m{(0,0)}}
		{\sum_{\chi,\chi'} \gamma^{\chi\chi'(2,0)}_{\bar{r}r}P_{\chi'}^\m{(0,0)}} \, ,
\end{equation}
with $P_\chi^\m{(0,0)}$ being the probability to find the dot in state $\chi$  and 
$\gamma^{\chi\chi'(2,0)}_{\bar{r}r}$ the transition rate from initial dot state $\chi'$ to final dot state $\chi$ where an electron is transferred from lead $r$ to lead $\bar{r}$. 
In the numerator,
only rates are taken into account that do not change the dot state.
In particular, no spin-flip processes are included. 
This contrasts with the expression in the denominator, in which spin-flip processes, i.e. $\chi=\sigma$ and $\chi'=\bar\sigma$ are taken into account.

In the limit of vanishing Coulomb interaction, $U=0$ (case 1), no spin-flip processes occur, which yields $c=1$.
Now we consider the limit of strong Coulomb interaction, $U=\infty$.
If the dot's level lies well above the Fermi energies of the leads, $\e \gg |\m{e}V/2|$ (case 2a), then the dot will be predominately empty.
Electrons passing through the quantum dot cannot flip their spin, and therefore $c=1$.
The situation becomes different for $-\e \gg |\m{e}V/2|$.
Then the dot is mostly singly occupied with either spin with probabilities $p_\uparrow$ and $p_\downarrow$ (such that $p_\uparrow^\m{(0,0)}+p_\downarrow^\m{(0,0)}=1$).
If the electrons travel from the ferromagnet to the normal lead (case 2b) we get $c=(1+p^2)/2$.
In the opposite case (case 2c), transport from the normal lead to the ferromagnet, we get always $c=1/2$ since an electron enters the dot from the normal lead and hence carries in one half of the cases the same spin as the electron initially occupying the dot.

These results are given in the last column of Table \ref{summary}.
The remaining question now is whether and how they are reflected in the visibility of the Aharonov-Bohm oscillations.

\subsection{Visibility}
 
\label{secfluxdep}
In the regime of unidirectional cotunneling, the leading order of transport through the quantum dot is $I^{(2,0)}$. In this limit the transmission through the QD is for all energies in good approximation the same. Furthermore, the $\sin \varphi$ part of the current which describes cotunneling through the lead but not through the quantum dot \cite{urban08} vanishes. The total current then is  
\begin{equation}
I^\m{total}=I^\m{(2,0)}+I^\m{(0,2)}+I^\m{(1,1)}=I^\m{av}\left(1+v \cos\varphi\right)\, ,
\end{equation} 
where $v$ is the visibility and $I^\m{av}$ the flux averaged current.

For the average current, measured in units of $I_0=\m{e}^2V/\pi$ we find
\begin{equation}
 \frac{I^\m{av}}{I_0}= \left\{ 
 	\begin{array}{ll}
	\displaystyle |t^\m{ref}|^2+\frac{\Gamma_\m{F}\Gamma_\m{N}}{\e^2} 
	& \text{for cases 1,2a,2b}
	\\
	\displaystyle  |t^\m{ref}|^2+(1-p^2)\frac{\Gamma_\m{F}\Gamma_\m{N}}{\e^2} 
	& \text{for case 2c} 
	\end{array} \right. 
\, .
\end{equation}

We express the visibility in terms of
\begin{equation}
v_0 = \frac{2\frac{\sqrt{\Gamma_\m{F}\Gamma_\m{N}}}{|\e|}|t^\m{ref}|}{|t^\m{ref}|^2+\frac{\Gamma_\m{F}\Gamma_\m{N}}{\e^2}} \; .
\end{equation}
We obtain 
\begin{equation}
 \frac{v}{v_0} = \left\{ 
 	\begin{array}{ll} 
	 \displaystyle 1
	 & \text{for cases 1,2a}
	 \\
	 \displaystyle  \frac{1+p^2}{2} 
	& \text{for case 2b}
	\\
	 \displaystyle \frac{1}{2} - \frac{p^2 |t^\m{ref}|^2}{2|t^\m{ref}|^2+
	 	2(1-p^2)\frac{\Gamma_\m{F}\Gamma_\m{N}}{\e^2}}
	& \text{for case 2c}
	\end{array} \right. 
\, .
\end{equation}

In the cases 1, 2a, and 2b, the visibility can be maximized by tuning $|t^\m{ref}|$ to $\frac{\sqrt{\Gamma_\m{F}\Gamma_\m{N}}}{|\e|}$.
Then, $v_0=1$ and $v=v_\m{max}$, independent of the degree of spin polarization $p$. 
However, for the case 2c, the maximal visibility $v_\m{max}$ can only be obtained by tuning $|t^\m{ref}|$ in a $p$-dependent way to $\frac{\sqrt{\Gamma_\m{F}\Gamma_\m{N}}}{|\e|}\sqrt{1-p^2}$. In this case $v_\m{max}$ turns out to be $v_\m{max}=\frac{\sqrt{1-p^2}}{2}$.

The visibility of the total current is a quantity that can be measured in an experiment. To what extent the visibility provides information about coherence of transport is discussed in the next section.

\subsection{visibility versus coherence}
\label{seccoherence}
\begin{figure}[h]
\centering
\includegraphics[width=6cm]{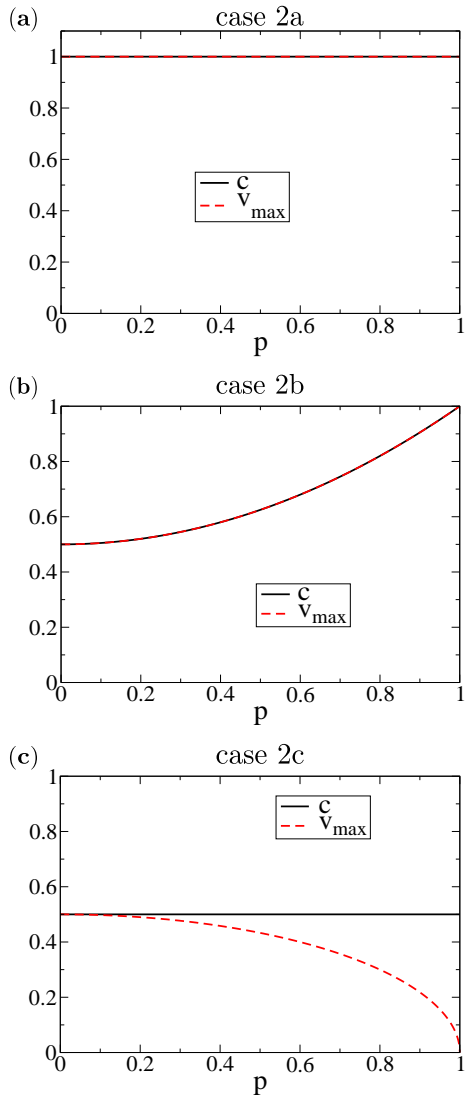}
\caption{\label{coherence} (Color online) Maximal visibility and coherent fraction of the total current for the two different transport directions.
} 
\end{figure}

In order to investigate the measurability of coherence we compare the fraction $c$ with the maximal visibility $v_\m{max}$. Because coherence is a essential assumption for flux dependence in general $c\ge v_\m{max}$. In the case of a vanishing Coulomb interaction (case 1) or a very high dot's level position (case 2a)  $v_\m{max}=1$ and, hence, $c=1$, see Fig.\ref{coherence}a). If the dot level is very low and electrons are transferred from the ferromagnet into the normal conductor (case 2b) the coherent fraction $c$ is equal to the maximal visibility $v_\m{max}$, see Fig. \ref{coherence}b). 
For a vanishing polarization $1/2$ of the electrons leaving the source carry the same spin as the electron initially occupying the dot. Hence, in one half of the cases the spin on the dot is not flipped and transport is coherent. The higher the polarization the more electrons with majority spin take part in transport and, thus, less spin-flip processes take place.\\	
For reversed transport voltages (case 2c) independent of the polarization one half of the processes are coherent. The source is a normal lead and, hence, one half of the electrons which tunnel onto the dot carry the same spin as the electron which initially occupied the dot. On the other hand the visibility is low for a high polarization, see Fig. \ref{coherence}(c), due to spin blockade on the dot. While transport through the reference arm is spin independent transport through the quantum dot is not. This prevents the possibility to tune the transmission through the reference arm and the transmission through the quantum dot to the same value.\\
In all cases the maximal visibility is obtained by tuning $|t^\m{ref}|$ to a certain value. While in the cases 1, 2a and 2b this value is independent of the polarization $p$, in case 2c $|t^\m{ref}|$ has to be tuned in a $p$-dependent way, see Sec. \ref{secfluxdep}.\\

\begin{widetext}
\begin{center}
\begin{table}[h]\caption{\label{summary}Summary of results}
\begin{tabular}{|c|c|c|c|}
\hline
&$v/v_0$&$v_\m{max}$&c\\ \hline\hline
$U=0$ (case 1) &$1$&1&1
\\ \hline
$U=\infty$, $\e\gg|\m{e}V/2|\gg k_\m{B}T$ (case 2a) &$1$&1&1
\\ \hline
$U=\infty$, $-\e\gg|\m{e}V/2|\gg k_\m{B}T$,$F\rightarrow N$ (case 2b) &$(1+p^2)/2$&$(1+p^2)/2$&$(1+p^2)/2$
\\ \hline
$U=\infty$, $-\e\gg|\m{e}V/2|\gg k_\m{B}T$,$N\rightarrow F$ (case 2c)&$\frac{1}{2} - \frac{p^2 |t^\m{ref}|^2}{2|t^\m{ref}|^2+2
	 	(1-p^2)\frac{\Gamma_\m{F}\Gamma_\m{N}}{\e^2}}$&$\sqrt{1-p^2}/2$&1/2
\\ \hline
\end{tabular}
\end{table}
\end{center}
\end{widetext}

\section{Conclusions}
\label{conclusions}
We have investigated the current through an AB-interferometer coupled to one normal and one ferromagnetic lead with a quantum dot embedded in one of the arms. In particular we elucidated the influence of polarization on the visibility of transport and studied the relation between visibility and coherence. We found that in the lowest flux-dependent order transport of noninteracting electrons is fully coherent and the maximal visibility is $1$. In the case of an infinite intra-dot Coulomb repulsion the coherence as well as the visibility of the current are strongly influenced by the polarization and the transport direction. As long as no spin blockade on the dot occurs the maximal visibility is equal to the coherent fraction of the current and can be obtained by tuning the transmission through the reference arm $|t^\m{ref}|$ in a $p$-independent way.

\acknowledgements
We acknowledge stimulating discussions with Saskia Fischer and Sven Buchholz, as well as support from DFG via SPP 1285, SFB 491 and EU STREP GEOMDISS. 

\appendix
\section{Calculation of a cotunneling rate}
\label{secordcalc}
In this appendix we give an example of how to calculate the cotunneling rates. For this, we choose the rate 
$\gamma^{\bar{\sigma}\sigma(2,0)}_{rr'}$, where an electron is transfered from lead $r'$ to lead $r$ accompanied by a change of the dot state from $\sigma$ to $\bar{\sigma}$. We start with
\begin{align}
\label{apprate1}
\gamma^{\bar{\sigma}\sigma(2,0)}_{rr'}&=\frac{1}{2\pi}\int\limits_{-\infty}^{\infty}d\omega f(\omega-\mu_{r'})\left(1-f(\omega-\mu_r)\right) \nonumber\\
&\times {\rm Re}\left[\frac{\sqrt{\Gamma_{r'}^{\bar{\sigma}}\Gamma_r^\sigma}}{\e-\omega+i0^+}+\frac{\sqrt{\Gamma_{r'}^{\bar{\sigma}}\Gamma_r^{\sigma}}}{\e+U-\omega+i0^+}\right]^2\, .
\end{align}
The regularization $+i0^+$ in the resolvents is added here by hand; however, it appears naturally when employing the real-time diagrams.
In the present case, there are two possible intermediate states, the dot being empty or doubly occupied. The corresponding energy differences to the final state appear in the denominator of the resolvents that have to be added coherently before performing the square.
The Fermi functions guarantee that the lead state from the electron enters the dot is occupied and that the lead state to which the dot electron leaves is unoccupied. The integral sums over all possible energies of the incoming electron.   For an infinite Coulomb repulsion $U=\infty$ and $|\e|\gg \max\{k_\m{B}T,\mu_r\}$ the rate simplifies to
\begin{align}
\gamma^{\bar{\sigma}\sigma(2,0)}_{rr'}&=\frac{1}{2\pi\e^2}\Gamma_{r'}^{\bar{\sigma}}\Gamma_r^\sigma\int\limits_{-\infty}^{\infty}d\omega f(\omega-\mu_{r'})\left(1-f(\omega-\mu_r)\right)\nonumber\\ 
&=\frac{1}{2\pi\e^2}\Gamma_{r'}^{\bar{\sigma}}\Gamma_r^\sigma\frac{\mu_r-\mu_{r'}}{e^{\beta(\mu_r-\mu_{r'})}-1}\, .
\end{align}
The calculation of the current rate is similar. The only difference is that one has to multiply the rates with the charge that is being transfered through the dot during the cotunneling process.

\end{document}